\documentclass[aps,pre,superscriptaddress,showpacs,showkeys,nofootinbib]{revtex4-1}

\usepackage{amssymb,amsmath,amsfonts}
\usepackage{graphicx}
\usepackage{color}
\usepackage{url}

\begin{document}
\title{Fokker-Planck Description of Wealth Dynamics and the Origin of Pareto's Law
}
\thanks{\copyright 2014, all rights reserved}
\author{Bruce M. Boghosian\footnote{
Present address:  College of Science and Engineering, American Univeristy of Armenia, Yerevan 0019, Republic of Armenia.}}
\affiliation{Department of Mathematics, Bromfield-Pearson Hall, Tufts University, Medford, Massachusetts 02155, USA\\
{\tt bruce.boghosian$@$tufts.edu}}
\date{January 31, 2014}
\begin{abstract}
The so-called ``Yard-Sale Model'' of wealth distribution posits that wealth is transferred between economic agents as a result of transactions whose size is proportional to the wealth of the less wealthy agent.  In recent work [B.M. Boghosian, ``Kinetics of Wealth and the Pareto Law,'' {\it Phys. Rev. E} {\bf 89} (2014) 042804], it was shown that this results in a Fokker-Planck equation governing the distribution of wealth.  With the addition of a mechanism for wealth redistribution, it was further shown that this model results in stationary wealth distributions that are very similar in form to Pareto's well known law.  In this paper, a much simpler derivation of that Fokker-Planck equation is presented.
\end{abstract}
\maketitle

\section{Introduction}

A key goal of modern economic theory is to explain how macroeconomic trends arise from microeconomic interactions.  In economic parlance, such explanations are known as {\it microfoundations}.  In recent years methods of the branch of statistical physics known as kinetic theory have been employed for this purpose, resulting in the field of {\it econophysics}.

One key empirical macroeconomic observation that has defied microfoundations for over a century is Pareto's celebrated law of wealth distribution~\cite{2}.  This states that the fraction of the population whose wealth exceeds $w$ is a power law in $w$ above a certain lower threshold.  To quantify this, suppose that wealth is distributed according to a probability density function (PDF) $P(w)$, so that $\int_a^b dw\; P(w)$ is the total population of economic agents with wealth $w\in[a,b]$.  The fraction of the population with wealth greater than $w$ is then
\begin{equation}
A(w) := \frac{1}{N}\int_w^{\infty} dw'\;P(w'),
\label{eq:ParetoA0}
\end{equation}
where $N:=\int_0^\infty dw\;P(w)$ is the total number of economic agents.  Pareto studied this function and found that it was well approximated by
\begin{equation}
A_p(w) =
\left\{
\begin{array}{ll}
1 & \mbox{if $w<w_{\min}$}\\
\left(\frac{w_{\min}}{w}\right)^{\alpha} & \mbox{otherwise,}
\end{array}
\right.
\label{eq:ParetoA1}
\end{equation}
where $w_{\min}$ is a lower threshold on wealth, and the exponent $\alpha$ is called the {\it Pareto index}.  Pareto indices are regularly measured by economists~\cite{3}, and tabulated for various countries in various epochs.

How might kinetic theory be brought to bear on the problem of explaining Pareto's law?  Kinetic theory usually deals with gases consisting of molecules, each endowed with spatial position and momentum, which undergo pairwise collisions that conserve momentum and energy, resulting in a Maxwell-Boltzmann distribution of velocities.  Ideally, econophysics ought to deal with economies consisting of agents, each endowed with wealth, who undergo pairwise transactions that conserve agents and wealth, resulting in a Pareto distribution of wealth.

A step toward making the compelling metaphor presented above rigorous was recently achieved by the demonstration~\cite{1} that a simple pairwise interaction between economic agents results in a Fokker-Planck equation for $P(w)$, whose steady-state solution in the presence of some amount of wealth redistribution is very similar in form to the Pareto distribution, Eq.~(\ref{eq:ParetoA1}).  In this paper, we re-derive this Fokker-Planck equation in a much more simple fashion than the original reference.

Section~\ref{sec:ys} describes the pairwise interaction from which the Fokker-Planck equation was derived, as well as a simple mechanism for wealth redistribution that was introduced in the earlier derivation~\cite{1}.  Both of these features are cast into the Fokker-Planck formalism by a simple derivation in Section~\ref{sec:fp}.  Section~\ref{sec:dc} contains discussion and conclusions.

\section{Asset-Exchange Model for the evolution of wealth distribution}
\label{sec:ys}

\subsection{Valuation}

We suppose that we have a closed population of $N$ agents that engage in pairwise transactions.  In these transactions, they exchange commodities, which we suppose have some intrinsic value.  The wealth of an agent is then the sum of the intrinsic values of all the commodities held by that agent.

When two agents exchange a commodity, they may do so for a price that is different from the intrinsic value.  This difference may be due to the agents' own personal utility functions -- that is, their own senses of the value of the commodity.  When this difference is nonzero, intrinsic wealth changes hands.

Before going on, it is necessary to acknowledge that the above description of wealth is somewhat of an anathema to neoclassical economists.  Most microeconomics texts define the value of a commodity as that which a buyer is willing to pay for it, and not as an intrinsic property of the commodity.

In reality, it is possible that one person will overvalue a commodity and another will undervalue it, and so both of them might be perfectly satisfied by an exchange at a price that is different from the intrinsic value.  The position we take in this paper, however, is that one of those two agents has gained some intrinsic wealth and the other has lost some.  In this case, we say that the second has made a {\it mistake}.

\subsection{Example:  The Norse Vikings}

To understand why this position is not as unreasonable as it might at first seem to an orthodox economist, consider the example of the Norse Vikings who colonized Greenland in AD 980, as described by Jared Diamond in his acclaimed book {\it Collapse: How Societies Choose to Fail or Succeed}~\cite{4}.  According to Diamond, the Vikings refused to eat fish which swam in abundance in Greenland's rivers.  At great expense, they also imported granite blocks from Norway so that they could build churches in Greenland that were similar to those they had left behind.  As a result of disastrous decisions such as these and many others, the colony collapsed in just under four hundred years.  By the end of the fifteenth century, from a maximum population of five thousand, there was not a single survivor.

While an economist might prefer to say that the Norse Vikings had a utility function that valued granite and did not value fish, the consequence of this perverse system of valuation was their ultimate destruction.  Indeed, Diamond's entire thesis is that societal choices may lead to both good and bad outcomes.  Some individuals and societies make good choices and others make bad ones, and often this determines their fate.  In this work, we take the view that fish had a higher {\it intrinsic} value than the Vikings accorded it, and granite had a lower {\it intrinsic} value.  From our perspective then, the Vikings' refusal to attribute an adequately high price to the former, and their insistence on attributing a very high price to the latter, were therefore {\it mistakes} which led to a loss of intrinsic wealth in the colony, ultimately weakening it.

Of course, a utility function that valued longevity (both individual and societal) in addition to personal preference might be able to capture the folly of the Vikings' position described above, but the point is that the Vikings did not behave in a manner designed to maximize such an improved utility function.  In fact, it is most likely that they did not even consciously recognize that they were making a harmful choice in this regard.  (One can safely assume that they were not running linear programming algorithms to make their decisions.)

Neoclassical economics assumes, inter alia, that (i) economic agents always behave rationally, and (ii) they always act to maximize their own profit.  If Diamond is correct, the Norse Vikings behaved contrary to both of these assumptions.  Their refusal to eat fish, for example, as evidenced by the lack of fish bones in their garbage middens, may have been due to a cultural or religious prohibition.  Such prohibitions, by their very nature, run counter to pure economic self interest, but they are often very difficult for people to disobey.  In this case, abandoning such prohibitions would have been better both for the individual and the collective, but the colony was apparently unable to summon the will to do so.

\subsection{The Yard-Sale Model}

More generally, when two economic agents engage in a transaction, one may make a mistake resulting in the transfer of wealth to the other.  According to the so-called {\it Yard-Sale Model}~\cite{5}, the amount of wealth transferred should be a specified fraction $\beta$ of the wealth of the less wealthy of the two agents, and the beneficiary of this excess goes to one agent or the other with even odds.  This assumption is reasonable because most people are unable to enter into transactions in which a large fraction of their own wealth is at stake, much less a large fraction of somebody else's wealth.  Hence, in what follows, we treat $\beta$ as small.

To quantify this, suppose that an agent with wealth $w$ transacts with another agent with wealth $w'$.  The total amount of wealth that is transferred from the second to the first in this transaction is then $\Delta(w,w',r) = \beta r\min\left( w,w'\right)$, where $r$ is a random variable equal to $+1$ or $-1$ with equal probability.  This may be rewritten
\begin{equation}
\Delta(w,w',r) = \beta r \left[w\theta\left(w'-w\right) + w'\theta\left(w-w'\right)\right],
\end{equation}
where $\theta$ denotes the Heaviside function
\begin{equation}
\theta(x) :=
\left\{
\begin{array}{rl}
1 & \mbox{if $x\geq 0$}\\
0 & \mbox{otherwise.}
\end{array}
\right.
\end{equation}
After the transaction, the wealth of the agents is
\begin{eqnarray}
w_{\mbox{\tiny final}} &=& w + \Delta(w,w',r)\\
w_{\mbox{\tiny final}}' &=& w' - \Delta(w,w',r).
\end{eqnarray}

\section{Derivation of the Fokker-Planck equation}
\label{sec:fp}

When a quantity such as $w$ varies in small stochastic steps $\Delta(w)$, its PDF, $P(w)$, is well known to obey a Fokker-Planck equation~\cite{6},
\begin{equation}
\frac{\partial P}{\partial t} =
-\frac{\partial}{\partial w}\left[\left\langle\Delta(w)\right\rangle P\right]
+ \frac{\partial^2}{\partial w^2}\left[\frac{\left\langle\Delta(w)\Delta(w)\right\rangle}{2} P\right],
\end{equation}
where the angle brackets $\left\langle\cdot\right\rangle$ denote the average over the distribution of step sizes.  The first term is often called the {\it drift} term, while the second is the {\it diffusive} term.

\subsection{Derivation of the Fokker-Planck equation}

The post-transaction wealth of an agent with wealth $w$ has two sources of stochasticity.  First, the wealth $w'$ of the agent's transaction partner needs to be sampled from the PDF $P(w')$.  Second, there is the random variable $r$.  We define the average of a function $f(w',r)$ over these two sources of stochasticity as follows
\begin{equation}
\left\langle f \right\rangle =
\frac{1}{N}\int_0^\infty dw'\; P(w') \frac{f(w',-1) + f(w',+1)}{2}.
\end{equation}

We can now take the average of the step size and its square as follows,
\begin{equation}
\left\langle\Delta(w)\right\rangle
=
\frac{1}{N}\int_0^\infty dw'\; P(w') \frac{\Delta(w,w',-1) + \Delta(w,w',+1)}{2}
= 0,
\end{equation}
and
\begin{eqnarray}
\left\langle\Delta(w)\Delta(w)\right\rangle
&=&
\frac{1}{N}\int_0^\infty dw'\; P(w') \frac{\left[\Delta(w,w',-1)\right]^2 + \left[\Delta(w,w',+1)\right]^2}{2}\\
&=&
\frac{\beta^2}{N} \int_0^\infty dw'\; P(w') \left[w\theta\left(w'-w\right) + w'\theta\left(w-w'\right)\right]^2\\
&=&
\frac{\beta^2}{N} \int_0^\infty dw'\; P(w') \left[w^2\theta\left(w'-w\right) + w'^2\theta\left(w-w'\right)\right]\\
&=&
\frac{\beta^2}{N} \int_w^\infty dw'\; P(w') w^2 + \frac{\beta^2}{N} \int_0^w dw'\; P(w') {w'}^2\\
&=&
2\beta^2\left[\frac{w^2}{2} A(w) + B(w)\right],
\end{eqnarray}
where $A(w)$ is Pareto's function defined in Eq.~(\ref{eq:ParetoA0}), and where we have defined the incomplete second moment
\begin{equation}
B(w) := \frac{1}{N} \int_0^w dw'\; P(w') \frac{{w'}^2}{2}.
\end{equation}

The Fokker-Planck equation describing the PDF is then
\begin{equation}
\frac{\partial P}{\partial t} = \beta^2\frac{\partial^2}{\partial w^2}\left[
\left(\frac{w^2}{2}A+B\right)P
\right],
\label{eq:ffpp}
\end{equation}
which is identical to the evolution equation discovered in earlier work~\cite{1} by a more general but much more laborious method of derivation.  By taking the zeroth and first moments of this dynamical equation, it is easily demonstrated that it conserves both $N$ and $W$.

\subsection{Redistribution}

In addition to the Yard-Sale Model described above, let us suppose that in a small time increment, an agent of wealth $w$ is taxed at rate $\tau$, thereby giving up wealth $\tau w$ to the tax collector.  The total tax collected from all agents is then $\int_0^\infty dw\; P(w)\tau w = \tau W$, where we have defined the total wealth as the first moment,
\begin{equation}
W = \int_0^\infty dw\; P(w) w.
\end{equation}
If we redistribute the total tax collected by dividing it equally amongst the $N$ agents, then each agent has gained the net amount of wealth
\begin{equation}
\Delta_r(w) = \tau\left(\frac{W}{N} - w\right),
\end{equation}
where the subscript ``r'' denotes ``redistribution.''  This will be positive for some agents and negative for others.  Note that it is not a stochastic variable, so it contributes to the drift term but not to the diffusive term.  Adding it into the drift term of our Fokker-Planck equation, we obtain
\begin{equation}
\frac{\partial P}{\partial t} =
-\tau\frac{\partial}{\partial w}\left[\left(\frac{W}{N} - w\right)P\right]
+\beta^2\frac{\partial^2}{\partial w^2}\left[
\left(\frac{w^2}{2}A+B\right)P
\right],
\end{equation}
We can absorb the factor of $\beta^2$ into the definition of the time scale, thereby effectively adopting units of time on the order of the time between transactions.  Defining the tax rate per transaction time $\chi=\tau/\beta^2$, we finally write
\begin{equation}
\frac{\partial P}{\partial t} =
-\chi\frac{\partial}{\partial w}\left[\left(\frac{W}{N} - w\right)P\right]
+\frac{\partial^2}{\partial w^2}\left[
\left(\frac{w^2}{2}A+B\right)P
\right],
\label{eq:ss}
\end{equation}
Once again, it is straightforward to show that this dynamical equation conserves both $N$ and $W$.

\subsection{Steady-state solutions}

To investigate steady-state solutions of Eq.~(\ref{eq:ss}), we can set the time derivative to zero and integrate once with respect to $w$ to obtain
\begin{equation}
\frac{\partial}{\partial w}\left[
\left(\frac{w^2}{2}A+B\right)P
\right]-\chi\left(\frac{W}{N} - w\right)P = 0.
\label{eq:sss}
\end{equation}
In the earlier work~\cite{1}, solutions to Eq.~(\ref{eq:sss}) were shown to exhibit power-law behavior at large $w$ and cutoff behavior at small $w$, completely consistent with Pareto's observations.

To see, for example, why a solution with a sharp cutoff at low $w$ is a solution, let us assume that $P(w)$ is small at low $w$ and justify that assumption a posteriori.  Small $P(w)$ at low $w$ means that $A(w)\approx 1$ and $B(w)\approx 0$ in this vicinity.  Under these circumstances, Eq.~(\ref{eq:sss}) reduces to
\begin{equation}
\frac{\partial}{\partial w}\left(
\frac{w^2}{2}P
\right)-\chi\left(\frac{W}{N} - w\right)P = 0,
\label{eq:ssss}
\end{equation}
which admits the solution
\begin{equation}
P(w) = \frac{C}{w^{2+2\chi}}\exp\left(-2\chi\frac{W}{N}\frac{1}{w}\right),
\label{eq:nonanalytic0}
\end{equation}
where $C$ is a constant of integration.  This function and all its derivatives vanish in the limit as $w\rightarrow 0$, and hence its Taylor series about $w=0$ vanishes.  On the other hand, the function itself is nonzero for $w>0$, and so it is non-analytic at $w=0$.  As a result, it is exceedingly small and flat for small $w$, providing us with the desired a posteriori justification.

For larger $w$, Eqs.~(\ref{eq:ssss}) and (\ref{eq:nonanalytic0}) are no longer valid because $A(w)$ is no longer near unity and $B(w)$ is no longer near zero.  In that regime, it is difficult to derive analytically that solutions to Eq.~(\ref{eq:sss}) increase sharply near a particular cutoff value, $w_{\min}$, and then fall off like a power law for larger $w$.  Nevertheless, numerical solutions to Eq.~(\ref{eq:sss}) presented in the earlier paper~\cite{1} demonstrate these features convincingly.

\section{Discussion and conclusions}
\label{sec:dc}

We have presented a far simpler derivation of the Fokker-Planck equation for wealth distribution, than that used in the first citation~\cite{1}.  We did this by noting that the Yard-Sale Model of wealth exchange is naturally a Fokker-Planck process, and therefore by computing the mean of the step size and its square.  We then noted that the redistribution mechanism also fit neatly into the Fokker-Planck framework.

This may be contrasted with the approach in the first citation~\cite{1}, which was to first derive the dynamical equation without assuming that $\beta$ is small.  This resulted in a Boltzmann equation that contained $\beta$ as a parameter.  Only afterward was the limit of small $\beta$ explored, resulting in Eq.~(\ref{eq:ffpp}).  The small $\beta$ limit is akin to the assumption, often made in plasma kinetic theory, of frequent small collisions, for which it is known that a Fokker-Planck equation results~\cite{7}.  Indeed, the drift and diffusive terms in this situation are also integrals over the PDF, called {\it Rosenbluth potentials}, and these are the analogs of our $A(w)$ and $B(w)$.

Pursuing this analogy further, it is interesting to contemplate other similarities between the Fokker-Planck equation of plasma physics and the one derived herein.  For example, in the presence of an imposed magnetic field and parallel electric field, the Fokker-Planck equation of plasma physics can lack a steady state due to the {\it runaway electron} problem.  As electrons pick up speed along the field, their propensity to scatter perpendicular to the field actually decreases, so ultimately they are undergoing free acceleration parallel to the field.  Likewise, if the tax rate is set to zero, $\chi=0$, our economic Fokker-Planck equation also lacks an equilibrium, instead converging to a generalized function in the sense of a certain Sobolev norm~\cite{1}, as one agent runs away with all the wealth.  Of course, there are differences between these phenomena:  The lack of steady state in the plasma is due to an exogenous factor, namely the imposed parallel electric field, whereas the lack of steady state in the economy is due to the lack of an exogenous factor, namely the absence of taxation.

In any case, it is hoped that the simplicity of this derivation will make the methodology and the result more accessible to researchers who wish to study modifications and embellishments to this general model.

\appendix




\begin{thebibliography}{0}
\bibitem{1} B.M. Boghosian, ``Kinetics of Wealth and the Pareto Law,'' {\it Phys. Rev. E} {\bf 89} (2014) 042804.
\bibitem{2} V. Pareto, ``The Mind and Society -- Trattato Di Sociologia Generale,'' Harcourt, Brace (1935).
\bibitem{3} See, e.g., F. Alvaredo, A.B. Atkinson, T. Piketty, E. Saez, ``The World Top Incomes Database,'' {\tt http://g-mond.parisschoolofeconomics.eu/topincomes}.
\bibitem{4} J. Diamond, ``Collapse: How Societies Choose to Fail or Succeed,'' Viking Penguin (2005).
\bibitem{5} B. Hayes, ``Follow the money,'' {\it American Scientist} {\bf 90} (2002) 400-405.
\bibitem{6} H. Risken, ``The FokkerÐPlanck Equation: Methods of Solutions and Applications,'' second edition, {\it Springer Series in Synergetics}, Springer, ISBN 3-540-61530-X (third printing, 1996).
\bibitem{7} M.N. Rosenbluth, W.M. MacDonald, D.L. Judd, ``Fokker-planck equation for an inverse-square force,'' {\it Phys. Rev.} {\bf 107} (Jul 1957) 1Ð6.
\end{thebibliography}
\end{document}